\documentclass{ws-ijmpe}
\usepackage[super,compress]{cite}
\usepackage{lineno}
\begin{document}

\markboth{J. Singh, M.~U.~Ashraf, A.~M.~Khan, and S.~Kabana}{Strangeness production in $O+O$ collisions}

\catchline{}{}{}{}{}

\title{Strangeness production in $O+O$ collisions at $\sqrt{s_{\mathrm{NN}}}=7$~TeV using various model approaches}

\author{J.~Singh\footnote{J.~Singh}}

\address{Instituto de Alta Investigaci\'on, Universidad de Tarapac\'a, Casilla 7D, Arica, 1000000, Chile\\ 
jsingh2@bnl.gov}

\author{M.~U.~Ashraf}
\address{Department of Physics and Astronomy, Wayne State University, 666 W. Hancock, Detroit, Michigan 48201, USA\\
}

\author{A.~M.~Khan}
\address{Georgia State University, Atlanta, GA 30303, USA\\
}

\author{S.~Kabana}
\address{Instituto de Alta Investigaci\'on, Universidad de Tarapac\'a, Casilla 7D, Arica, 1000000, Chile\\
}

\maketitle
\begin{history}
\received{Day Month Year}
\accepted{Day Month Year}
\end{history}

\begin{abstract}
We present the predictions of various observables for strange ($\mathrm{K}^{0}_{\mathrm S}$, $\Lambda$($\overline{\Lambda}$)) and multi-strange hadrons ($\Xi^-$($\overline{\Xi}^+$), $\phi$, and $\Omega^-$($\overline{\Omega}^+$)) using the recently updated 3+1D hydrodynamics-based EPOS4 framework and AMPT model. In this study, we report the transverse momentum (${p_{\rm T}}$) spectra, particle yields (${\mathrm{d}N/\mathrm{d}y}$), and ${p_{\rm T}}$ integrated yield ratios relative to pions for $O+O$ collisions at $\sqrt{s_{\mathrm{NN}}}=7$~TeV. The results reveal that there are indications of stronger radial flow in EPOS4 compared to AMPT. We observe a final state multiplicity overlap with small ($p+p$ and $p+Pb$) and large ($Pb+Pb$) collision systems. 
\end{abstract}


\ccode{PACS numbers: 25.75.-q, 25.75.Nq, 25.75.Ld}

\section{Introduction}
High-energy heavy-ion ($A+A$) collisions offer a unique opportunity to create and study the Quark-Gluon Plasma (QGP) -- a de-confined state of quarks and gluons that emerges under extreme temperature and/or energy density conditions~\cite{1,2,3}. Experiments at the Relativistic Heavy Ion Collider (RHIC) and the Large Hadron Collider (LHC) have facilitated investigations on the properties of the QGP~\cite{rhic1,lhc1}. Strangeness enhancement has been widely recognized and observed as a key signature of the QGP, as the hot medium facilitates the thermal production of strange quarks~\cite{4}. The yield ratios of various strange hadrons to pions in A+A collisions, from SPS to LHC, exhibit a strong dependence on centrality and collision energy~\cite{5}. Recent LHC studies~\cite{6, 7} have revealed unexpected similarities between high-multiplicity $p+p$ and $p+A$ and $A+A$ collisions, including azimuthal anisotropies, modified hadron yields, and spectral changes, despite the significant different system sizes~\cite{6,8}.

These findings challenge the current theoretical understanding of QGP formation, suggesting that QGP-like behavior may emerge in smaller collision systems than previously expected, with the notable exception of an early theoretical work that proposed QGP formation occurs once a threshold in the initial energy density is surpassed, regardless of the system size, whether in $p+p$, $p+A$, or $A+A$ collisions~\cite{kabanaminkowski}. To deepen our understanding of QGP formation in small systems, LHC experiments are anticipated to collect and analyze the data from oxygen-oxygen ($O+O$) collisions at $\sqrt{s_{\mathrm{NN}}}~\approx~7$ TeV~\cite{OO1,OO2,smallsystem}. This provides a crucial opportunity to study QGP-like effects in a system with a limited number of participating nucleons, a final-state multiplicity comparable to smaller systems, and a larger transverse overlap. Investigating strangeness production in $O+O$ collisions, which bridges $p+p$ and $p+Pb$ at low multiplicities and $Pb+Pb$ at higher multiplicities~\cite{OO2}, is essential for understanding the hadrochemistry of the medium and the process of hadronization.

The predictions presented in this work are based on recently updated EPOS framework (EPOS4)~\cite{EPOS41,EPOS42,EPOS43} and two different versions of AMPT model~\cite{AMPT1,AMPT2,transport}. Both models employ different mechanisms for strangeness production. EPOS4 includes the QGP phase in its hydrodynamic evolution with the lattice QCD equation of state (EoS). AMPT, on the other hand, does not have a fully chemically equilibrated system as compared to EPOS4. In this article, we report on the predictions of (multi-)strange hadron ($\mathrm{K}^{0}_{\mathrm S}$, $\Lambda+\overline{\Lambda}$, $\Xi^{-}+\overline{\Xi}^{+}$, $\Omega^- + \overline{\Omega}^+$, and $\phi$) production in $O+O$ collisions at $\sqrt{s_{\mathrm{NN}}}=7$~TeV using EPOS4 and AMPT. The observables under study include $p_{T}$ spectra, rapidity density distributions ($dN/dy$), and multiplicity dependence of the yield ratios of strange hadrons relative to pions.

\section{Event Generators}
\begin{romanlist}[(ii)]
\item {\bf EPOS4} framework is a multipurpose event generator which is used to simulate collisions across various systems, including $p+p$, $p+A$, and $A+A$ interactions~\cite{EPOS41,EPOS42,EPOS43}. It employs a 3+1D viscous hydrodynamic framework to model the evolution of $A+A$ collisions, with initial conditions based on flux tubes derived from Gribov-Regge multiple scattering theory~\cite{grtheory}. A distinctive feature of EPOS4 is its core-corona approach, which differentiates between dense core regions undergoing hydrodynamic evolution and less dense corona regions that hadronize into jets. The model also integrates the vHLLE algorithm for hydrodynamic evolution~\cite{vhhle1} and utilizes the UrQMD model for simulating the hadronic cascade in the later stages of collisions~\cite{urqmd1,urqmd2}.\\

\item A Multi Phase Transport ({\bf AMPT}) model is a transport model which is developed to investigate the dynamics of relativistic heavy-ion collisions and has been widely employed to study various observables at RHIC and LHC energies~\cite{AMPT1,AMPT2,transport}. In the AMPT model, HIJING  initializes the spatial and momentum distributions of minijet partons and soft string excitations~\cite{hijing1}. The subsequent space-time evolution of these partons is dictated by the ZPC parton cascade model~\cite{cascade1}. Following the partonic cascade, the model converts the remaining partonic degrees of freedom into final-state hadrons through either string fragmentation or quark coalescence. Subsequently, the interactions of these newly formed hadrons are governed by A Relativistic Transport (ART) model~\cite{art1}. This work utilizes both the default (AMPT-Def) and string melting (AMPT-SM) versions.

\end{romanlist}

A total of $\sim3$ million minimum-bias events were simulated using EPOS4, while $\sim6$ million minimum-bias events were generated using AMPT-SM and AMPT-Def, respectively. The multiplicity classes in these predictions are determined based on the pseudo-rapidity distribution of charged particles within $|\eta| < 0.5$, following the same methodology adapted by the experiments~\cite{centrality}.

\section{Results and Discussions}
\begin{figure}[th]
    \centering       
    \includegraphics[width=0.46\linewidth]{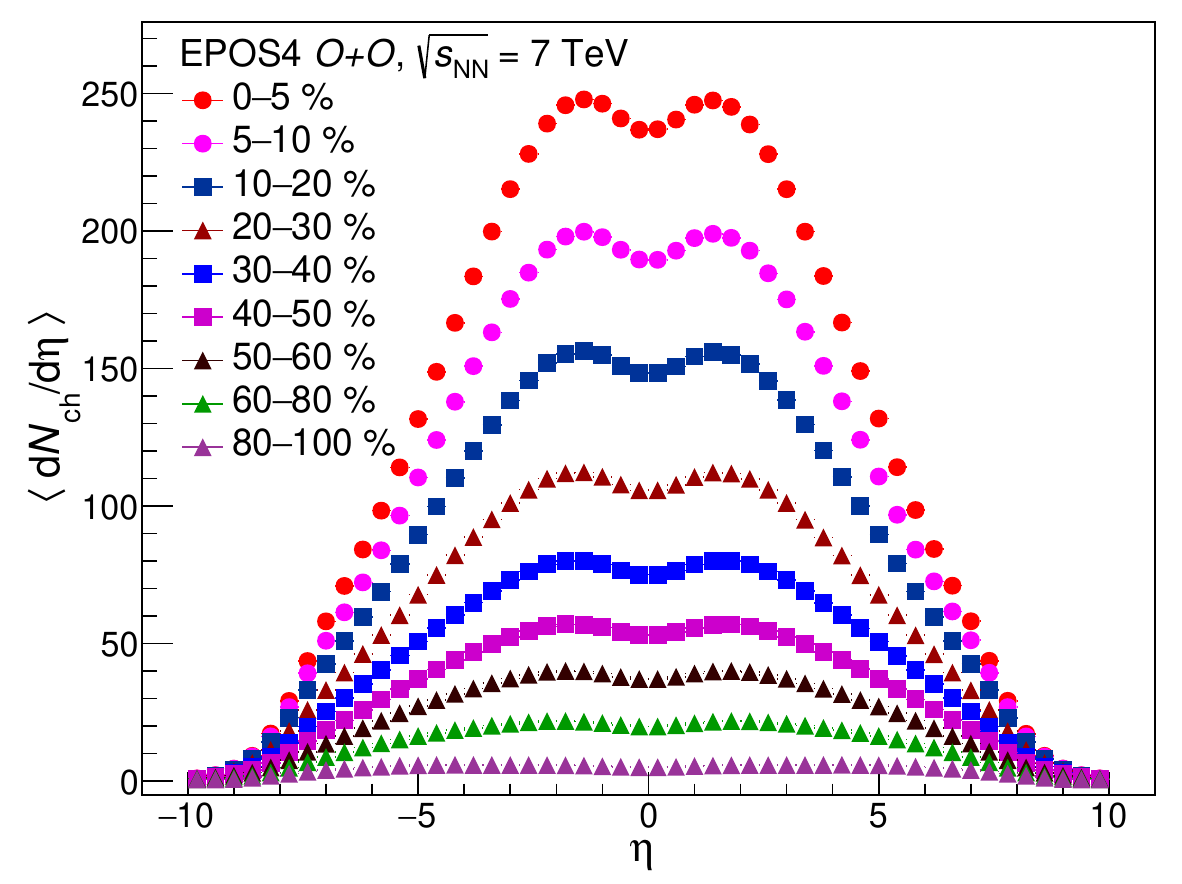} 
    \includegraphics[width=0.46\linewidth]{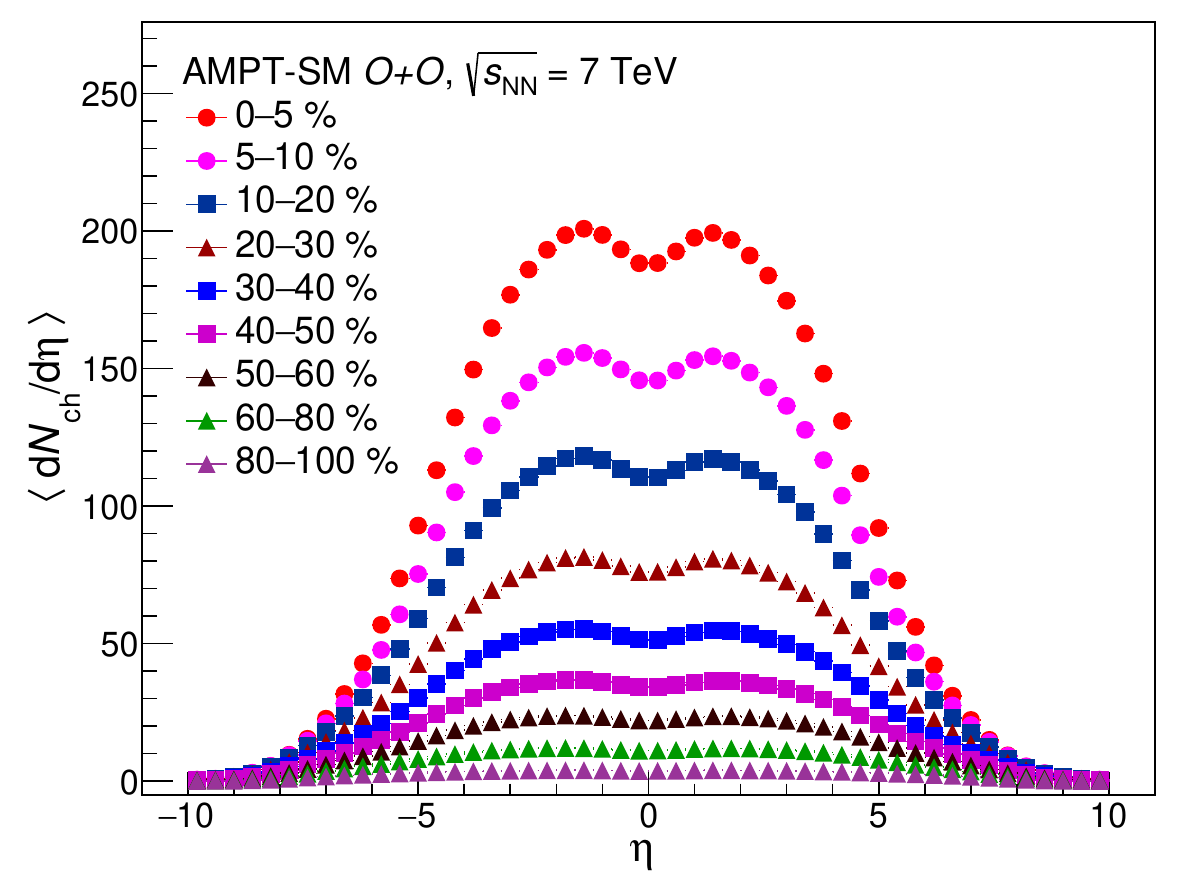}
    \caption{(Color online) Pseudo-rapidity distributions of charged particles in $O+O$ collisions at $\sqrt{s_{\mathrm{NN}}}=7$ TeV
for different centrality classes using EPOS4~\cite{Identified_OO,strange_OO7} and AMPT-SM model. Marker style with different colors represent different centrality classes.}\label{fig1}
\end{figure}

Figure~\ref{fig1} shows the charge particle multiplicity in $O+O$ collisions at $\sqrt{s_{\mathrm{NN}}}=7$ TeV for different centrality classes using EPOS4 and AMPT-SM. It is evident from Fig.~\ref{fig1} that EPOS4 predicts higher multiplicity as compared to AMPT-SM. Details of the $\langle dN_{ch}/d\eta \rangle$ and $\langle N_{\rm part} \rangle$ values across different centrality classes for EPOS4 and AMPT are provided in Ref.~\cite{Identified_OO,strange_OO7}. Figure~\ref{fig1} (right) is presented here for the first time.

\begin{figure}[th]
    \centering       
    \includegraphics[width=0.60\linewidth]{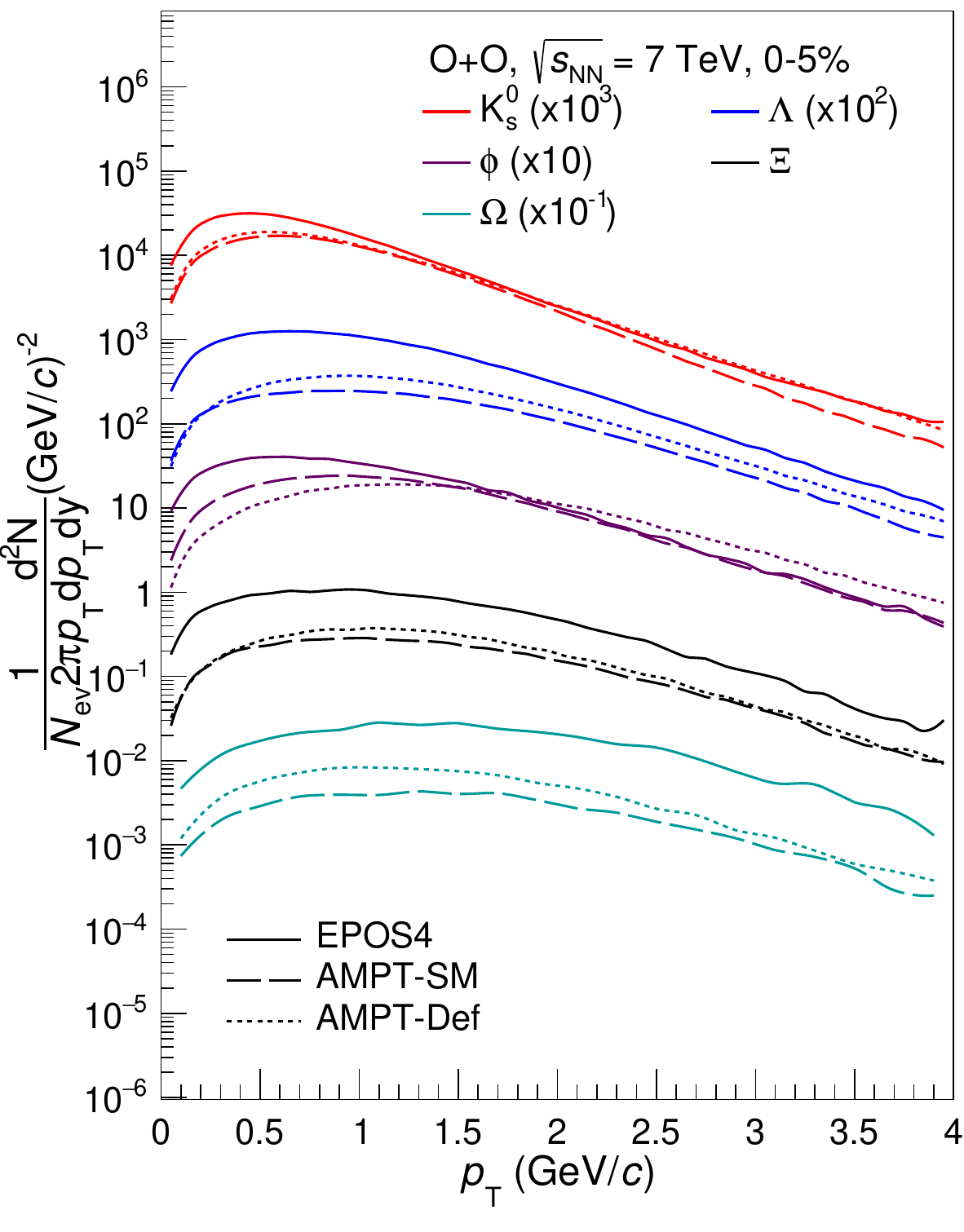}
    \caption{(Color online) $p_T$-spectra of (multi-)strange hadrons ($\mathrm{K}^{0}_{\mathrm S}$, $\Lambda+\overline{\Lambda}$, $\Xi^{-}+\overline{\Xi}^{+}$, $\Omega^- + \overline{\Omega}^+$, and $\phi$) from EPOS4 and  AMPT in $O+O$ collisions at $\sqrt{s_{\mathrm{NN}}}=7$~TeV for 0--5\% centrality classes. $p_T$ spectra for different hadrons are scaled by a factor of $10$ for better visualization. In figure, solid lines represent EPOS4, while dashed lines correspond to AMPT-SM, and dotted lines represent AMPT-Def.}
    \label{fig2}
\end{figure}

The transverse momentum ($p_{T}$) distributions of (multi-)strange ($\mathrm{K}^{0}_{\mathrm S}$, $\Lambda+\overline{\Lambda}$, $\Xi^{-}+\overline{\Xi}^{+}$, $\Omega^- + \overline{\Omega}^+$, and $\phi$) hadrons from AMPT-Def, AMPT-SM, and EPOS4 in the most central (0--5\%) $O+O$ collisions at $\sqrt{s_{\mathrm{NN}}}=7$~TeV are shown in Fig.~\ref{fig2}.
The $p_T$ distributions of various hadron species show a clear dependence on collision centrality and particle type.
At intermediate $p_T$, the spectra of heavier particles appear to converge with those of lighter particles, likely due to the effects of radial flow. Additionally, the $p_T$ spectra of heavier particles exhibit a significantly flatter slope compared to lighter particles, suggesting an increase in radial flow with rising particle mass. A similar trend has been observed in the $p_{T}$ spectra of identified hadrons in $O+O$ collisions using EPOS4~\cite{Identified_OO}. The production mechanisms of strange hadrons in AMPT and EPOS4 differ significantly, particularly at low $p_T$, where AMPT is driven by partonic scatterings and hadronic rescatterings, while EPOS4 is dominated by hydrodynamic evolution, leading to stronger radial flow.  In AMPT-SM, at intermediate $p_{T}$, quark coalescence plays an important role, while collective flow and fragmentation-recombination hadronization produces smoother spectra in EPOS4.

\begin{figure}[th]
    \centering       
    \includegraphics[width=0.55\linewidth]{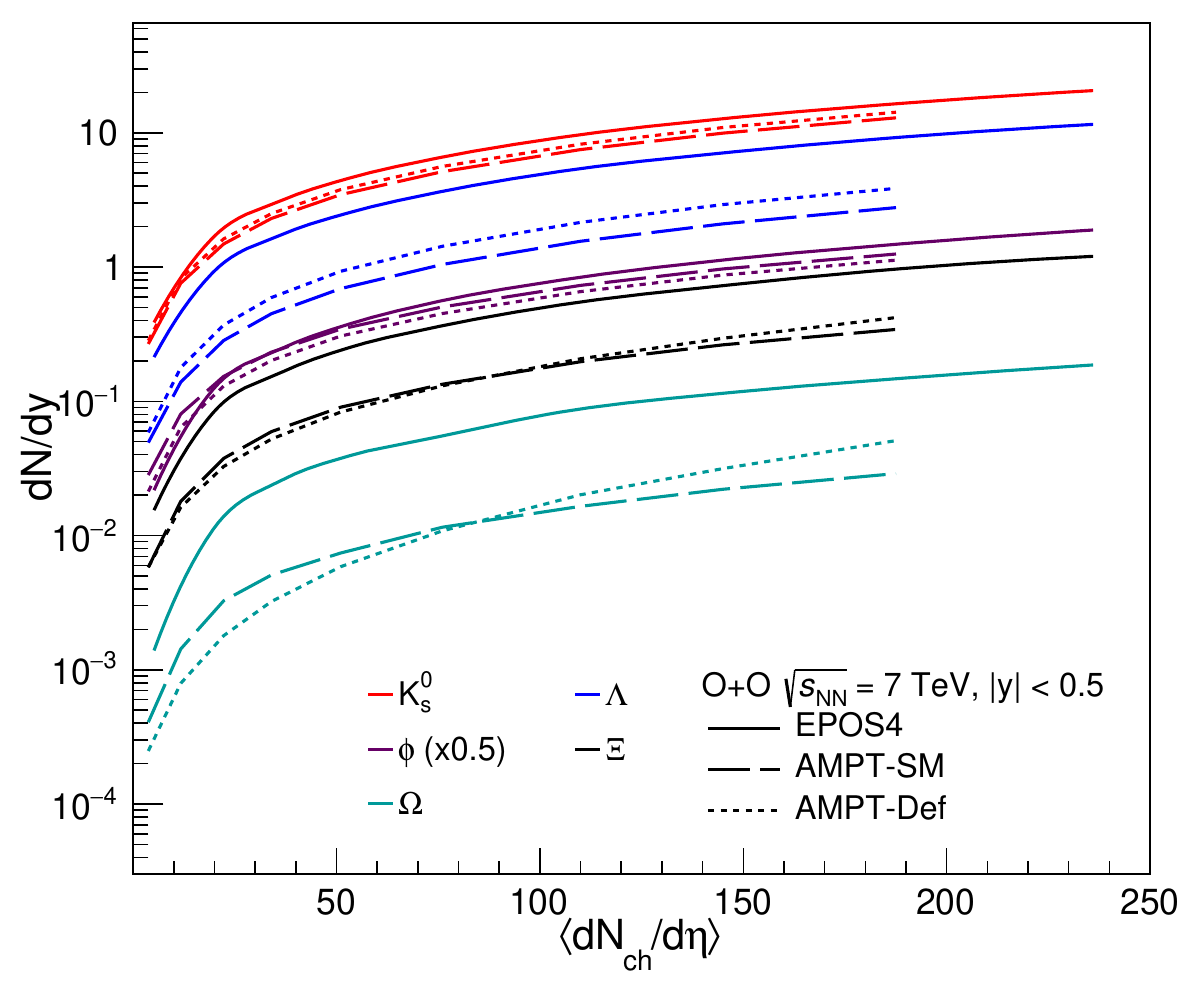} 
    \caption{(Color online) Multiplicity dependence of the integrated yield ($dN/dy$) at mid-rapidity $\mid y \mid< 0.5$ for (multi-)strange hadrons ($\mathrm{K}^{0}_{\mathrm S}$, $\Lambda$, $\Xi$, $\phi$, and $\Omega$) in $O+O$ collisions at $\sqrt{s_{\mathrm{NN}}}=7$ TeV using AMPT-Def, AMPT-SM and EPOS4.}\label{fig3}
\end{figure}
Figure~\ref{fig3} illustrates the integrated yields ($dN/dy$) of various strange hadrons as a function of charged-particle multiplicity ($\langle dN_{\text{ch}}/d\eta \rangle$) in $O+O$ collisions at $\sqrt{s_{\mathrm{NN}}}=7$~TeV at mid-rapidity ($\lvert y \rvert < 0.5$) for AMPT and EPOS4. Both models exhibit an increasing trend in the yields of strange hadrons from peripheral to central collisions, indicating a strong multiplicity dependence in peripheral collisions. 
Additionally, a systematic decrease in the integrated yields of multi-strange hyperons is observed with increasing strangeness content (i.e., number of constituent strange quarks).
All models consistently predict an increase in $dN/dy$ with increasing $\langle dN_{\text{ch}}/d\eta \rangle$. The EPOS4 predictions for $\Lambda$, $\Xi$, and $\Omega$ are systematically higher than those from AMPT-Def and AMPT-SM.

\begin{figure}[th]
    \centering       
    \includegraphics[width=0.60\linewidth]{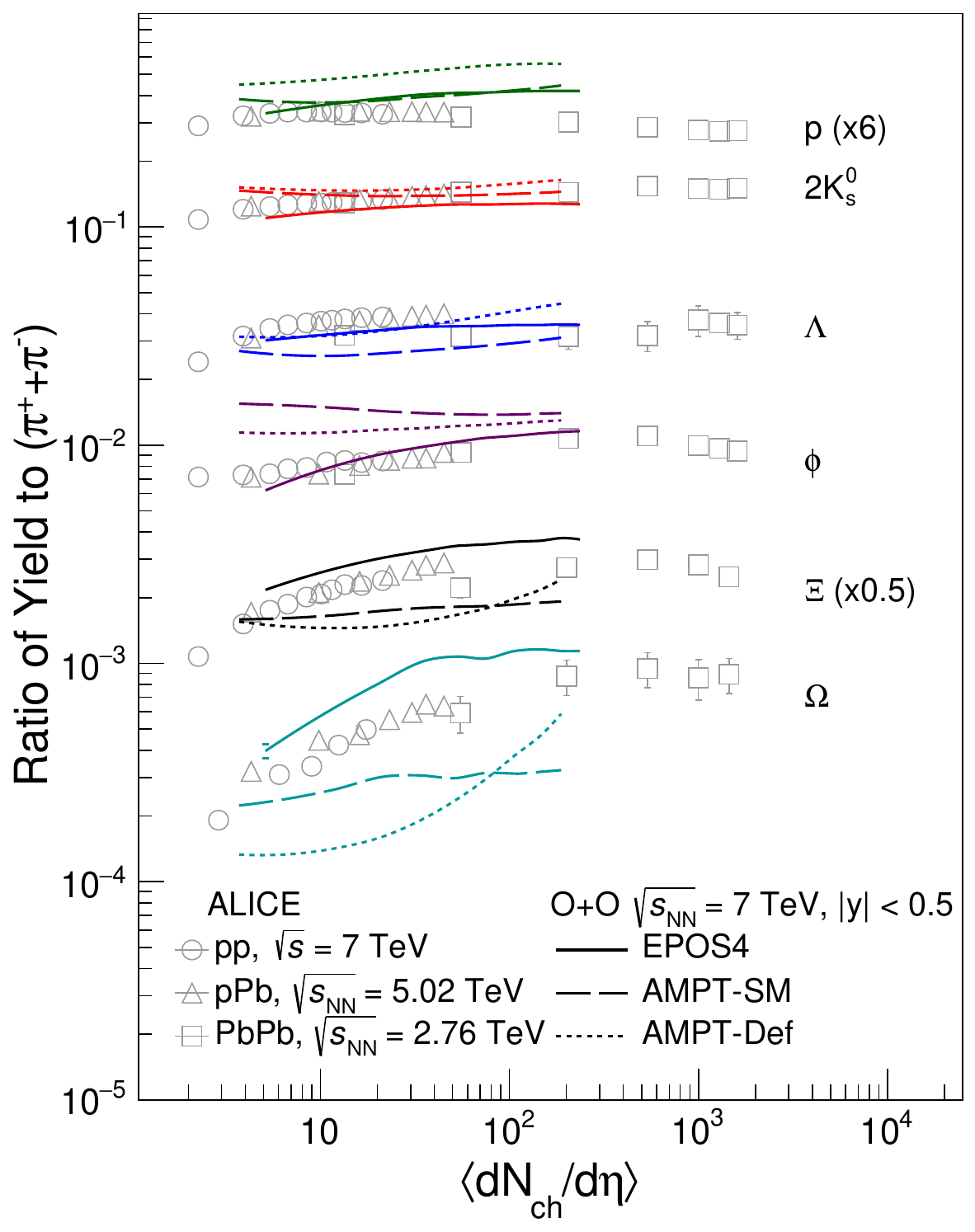} 
    \caption{(Color online) $p_T$-integrated yield ratios of $K_S^0$, $\Lambda$, $\Xi$, $\phi$, and $\Omega$ to pions ($\pi^+ + \pi^-$) as a function of $\langle dN_{\text{ch}}/d\eta \rangle$ in $O+O$ collisions at $\sqrt{s_{\mathrm{NN}}}=7$ TeV using AMPT-Def, AMPT-SM and EPOS4. Solid lines are used for EPOS4, whereas the dotted and dashed lines represent the AMPT-Def and AMPT-SM models, respectively. The values are compared
    to the published results from $p+p$, $p+Pb$, and $Pb+Pb$ collisions~\cite{alice_1,alice_2,alice_3,alice_4}.}\label{fig4}
\end{figure}

To examine the relative production of strange hadrons with respect to non-strange hadrons, the yield ratios of strange hadrons to pions were computed as a function of charged-particle multiplicity, as shown in Fig.~\ref{fig4}. In AMPT-Def, the $p_T$-integrated ratios exhibit weaker dependence on multiplicity, as strange quark production is primarily governed by fragmentation rather than collective effects. In contrast, the AMPT-SM model generates quarks through parton-parton scatterings, with hadronization occurring via quark coalescence. AMPT-SM is constructed in a way that provides a more realistic simulation of a dense medium similar to the QGP, leading to enhanced strangeness production. As charged-particle multiplicity rises, quarks in the AMPT-SM model have a higher probability of recombining into strange hadrons, resulting in an enhanced yield of $K_S^0$, $\Lambda$, $\Xi$, $\phi$, and $\Omega$ relative to pions. The enhancement of strange hadrons is more significant at higher multiplicities due to the favorable conditions for strange quark coalescence. Furthermore, AMPT simulations indicate that 0–5$\%$ centrality in O+O collisions corresponds to a charged-particle multiplicity comparable to that observed in 50–60$\%$ centrality in Pb+Pb collisions.

In EPOS4 framework, the collision dynamics use viscous hydrodynamics, capturing collective effects and medium evolution. At high multiplicities, the core region dominates, leading to enhanced strangeness production due to higher energy density. This results in a significant increase in strange hadron-to-pion ratios, particularly for multi-strange baryons like $\Xi$ and $\Omega$. In contrast, low-multiplicity events are corona-dominated, where hard processes suppress strangeness production. The predicted yield ratios of strange hadrons to pions in O+O collisions, obtained from all models, are compared with experimental measurements from $p+p$, $p+Pb$, and $Pb+Pb$ collisions at available LHC energies~\cite{alice_1,alice_2,alice_3,alice_4}. Notably, the model predictions exhibit a clear overlap in final-state multiplicity with these collision systems, in the $\langle dN_{\text{ch}}/d\eta \rangle$ range from $\sim$5 to $\sim$185 for AMPT  and $\sim$5 to $\sim$230 for EPOS4, suggesting a possible continuity in strangeness production mechanisms across different system sizes.

\section{Conclusions}
This study investigates predictions for transverse momentum ($p_T$) spectra, the multiplicity dependence of the particle yields ($\mathrm{d}N/\mathrm{d}y$), ${p_{\rm T}}$-integrated yield ratios relative to pions for (multi-)strange hadrons ($\mathrm{K}^{0}_{\mathrm S}$, $\Lambda$($\overline{\Lambda}$), $\Xi^-$($\overline{\Xi}^+$), $\phi$, and $\Omega^-$($\overline{\Omega}^+$)) in $O+O$ collisions at $\sqrt{s_{\mathrm{NN}}}=7$~TeV using the recently developed EPOS4 framework and the AMPT model. EPOS4 systematically predicts higher $\mathrm{d}N/\mathrm{d}y$ for $\Lambda$, $\Xi$, and $\Omega$ compared to AMPT-Def and AMPT-SM. EPOS4 performs well in predicting the strangeness enhancement, whereas the current version of AMPT exhibits significant limitations. None of the models can fully describe trends of strangeness enhancement observed in existing experimental data from other collision systems.
Interestingly, a final state multiplicity overlap  is observed in the yield ratios relative to pions when compared with published results of the small ($p+p$ and $p+Pb$) and large ($Pb+Pb$) systems. The upcoming data on $O+O$ collisions at $\sqrt{s_{\mathrm{NN}}}~\approx~7$~TeV at the LHC will provide crucial insights for constraining and refining model parameters.

\end{document}